\documentclass[numberedappendix]{emulateapj}
\usepackage{graphics}
\usepackage{amsmath}
\usepackage{amssymb}
\usepackage{natbib}

\usepackage[usenames,dvipsnames,svgnames]{xcolor}
\usepackage{graphicx}
\usepackage{verbatim}

\newcommand{\pc}{\mathrm{pc}}

\newcommand{\km}{\mathrm{km}}

\newcommand{\s}{\mathrm{s}}
\newcommand{\yr}{\mathrm{yr}}
\newcommand{\erg}{\mathrm{erg}}
\newcommand{\kms}{\mathrm{km~s}^{-1}}
\newcommand{\msun}{M_\odot}

\newcommand{\snr}{\mathrm{S/N~pixel}^{-1}}
\newcommand{\p}{\mathrm{p}}
\renewcommand{\d}{\mathrm{d}}
\newcommand{\edd}{\mathrm{Edd}}
\newcommand{\bol}{\mathrm{bol}}

\newcommand{\bh}{\mathrm{BH}}
\newcommand{\los}{\mathrm{LoS}}
\newcommand{\ang}{\mathrm{\AA}}
\renewcommand{\ion}[2]{{\rm#1}\;\textsc{\MakeLowercase{#2}}}
\newcommand{\hb}{\mathrm{H}\beta{}}

\begin{document}

\shorttitle{Binary BH detection}

\title{Searching for Binary Supermassive Black Holes via
  Variable Broad Emission Line Shifts: Low Binary Fraction}
%\title{Searching Binary Supermassive Black Hole by Broad
%  Emission Line Shift Detection: Low Binary Fraction}

\author{Lile Wang$^1$, Jenny E. Greene$^1$, Wenhua Ju$^1$,
  Roman R. Rafikov$^2$, John J. Ruan$^3$, 
  and Donald P. Schneider$^{4,5}$}

\footnotetext[1]{Princeton University Observatory,
  Princeton, NJ 08544}
  
\footnotetext[2]{Institute for Advanced Study, 
  Einstein Drive, Princeton, NJ 08540}

\footnotetext[3]{Department of Astronomy, 
  University of Washington, Box 351580, 
  Seattle, WA 98195}

\footnotetext[4]{Department of Astronomy and Astrophysics, 
  The Pennsylvania State University, 
  University Park, PA 16802}
  
\footnotetext[5]{Institute for Gravitation and the Cosmos, 
  The Pennsylvania State University, 
  University Park, PA 16802}

\begin{abstract}
  Supermassive black hole binaries (SMBHs) are expected to
  result from galaxy mergers, and thus are natural
  byproducts (and probes) of hierarchical structure
  formation in the Universe. They are also the primary
  expected source of low-frequency gravitational wave
  emission. We search for binary BHs using time-variable
  velocity shifts in broad \ion{Mg}{II} emission lines of
  quasars with multi-epoch observations. First, we inspect
  velocity shifts of the binary SMBH candidates identified
  in \citet{2013ApJ...777...44J}, using SDSS spectra with an
  additional epoch of data that lengthens the typical
  baseline to $\sim 10~\yr$. We find variations in the
  line-of-sight velocity shifts over 10 years that are
  comparable to the shifts observed over 1-2 years, ruling
  out the binary model for the bulk of our candidates. We
  then analyze 1438 objects with $8~\yr$ median time
  baselines, from which we would expect to see velocity
  shifts $>1000~\kms$ from sub-pc binaries. We find only one
  object with an outlying velocity of $448~\kms$,
  indicating, based on our modeling, {\bf that $\lesssim 1$
    per cent (the value varies with different assumptions)
    of SMBHs that are active as quasars reside in binaries}
  with $\sim 0.1~\pc$ separations.  Binaries either sweep
  through these small separations rapidly or stall at larger
  radii.
\end{abstract}
%%%%%%%%%%%%%%%%%%%%%%%%%%%%%%%%%%%%%%%%%%%%%%%%%%%%%%%%%%%%
% Abstract for plain text
%%%%%%%%%%%%%%%%%%%%%%%%%%%%%%
% Supermassive black hole binaries (SMBHs) are expected to
% result from galaxy mergers, and thus are natural
% byproducts (and probes) of hierarchical structure
% formation in the Universe. They are also the primary
% expected source of low-frequency gravitational wave
% emission. We search for binary BHs using time-variable
% velocity shifts in broad Mg II emission lines of quasars
% with multi-epoch observations. First, we inspect velocity
% shifts of the binary SMBH candidates identified in Ju et
% al. (2013), using SDSS spectra with an additional epoch of
% data that lengthens the typical baseline to ~10 yr. We
% find variations in the line-of-sight velocity shifts over
% 10 years that are comparable to the shifts observed over
% 1-2 years, ruling out the binary model for the bulk of our
% candidates. We then analyze 1438 objects with 8 yr median
% time baselines, from which we would expect to see velocity
% shifts >1000 km/s from sub-pc binaries. We find only one
% object with an outlying velocity of 448 km/s, indicating,
% based on our modeling, that ~< 1 per cent (the value
% varies with different assumptions) of SMBHs that are
% active as quasars reside in binaries with ~0.1 pc
% separations.  Binaries either sweep through these small
% separations rapidly or stall at larger radii.
%%%%%%%%%%%%%%%%%%%%%%%%%%%%%%%%%%%%%%%%%%%%%%%%%%%%%%%%%%%%


\section{Introduction}
\label{sec:introduction}

The mergers of supermassive black holes (SMBHs) likely play
a key role in SMBH formation and massive galaxy evolution
\citep[e.g.][] {1992ARA&A..30..705B, 2003ApJ...582..559V,
  2003ApJ...593..661V, 2005ApJ...620L..79S}. Since most
massive galaxies harbor SMBHs at their centers
\citep[see][and references therein]{2013ARA&A..51..511K},
and galaxies grow via hierarchical mergers, we expect binary
SMBHs to form, but we have no empirical constraints on the
rate of merging of these binary systems. As the most intense
sources of spacetime curvature, merging BHs are 
expected to be a dominant source of gravitational radiation
at low frequency \citep[e.g.][]{2011PhRvD..83l2005A,
  2013GWN.....6....4A}, dominating the low-frequency
gravitational wave background \citep[GWB; e.g.][]
{2001astro.ph..8028P, 2015MNRAS.447.2772R}. The merger rate
also has important implications for the spin distribution of
SMBHs \citep[e.g.][] {2009ApJ...704L..40B,
  2012arXiv1208.5251G, 2016ApJ...825L..19H}.

The typical SMBH merger scenario consists of several phases
of evolution \citep[e.g.][] {1980Natur.287..307B}. At large
radius (kpc scales), the SMBHs dissipate energy and angular
momentum by dynamical friction with stars in galactic
bulges, until the binary hardens and reaches an orbital
velocity that is greater than the typical velocity
dispersion of the bulge stars \citep[see e.g.][]
{1996NewA....1...35Q, 2005LRR.....8....8M}. At these $\sim$
parsec separations, in an axisymmetric system, there are no
longer sufficient stars to be scattered by the SMBH binary.
Gravitational wave emission cannot efficiently facilitate
the merger of BH binaries until $\lesssim 10^{-3}~\pc$
scales \citep[see also][]{1980Natur.287..307B}. In
principle, the SMBH binary may stall at $\sim 1~\pc$
indefinitely, a possibility known as the ``final parsec''
problem \citep[e.g.][]{1980Natur.287..307B}.  Many solutions
have been proposed to efficiently merge SMBH binaries within
a Hubble time, including a gas-assisted inspiral model
\citep[e.g.][] {1999MNRAS.307...79I, 2009ApJ...700.1952H,
  2013ApJ...774..144R, 2014PhRvD..90j4030G,
  2015A&A...576A..29I, 2016ApJ...827..111R}, or enhanced
efficiency of binary-star scattering in triaxial halos
\citep{2012ApJ...749..147K, 2015ApJ...810...49V,
  2016arXiv160400015K}. However, from both observational and
theoretical perspective, it remains unclear whether SMBH
binaries merge efficiently or stall forever.

The goal of the current work is to statistically constrain
the duration of the SMBH binary phase at $\sim 0.1$~pc
scales by looking through a large population of quasars with
multi-epoch spectroscopy for time-varying radial velocity
shifts caused by the binary orbital motion. A handful of
studies have attempted to image SMBH binaries using radio
observations \citep{2006ApJ...646...49R,
  2011MNRAS.410.2113B}, but in general it is prohibitive to
spatially resolve such binaries at cosmological redshifts
where the cosmic mass density in BHs is built up. Therefore,
time-domain searches are required. One approach is to use
light-curves to search for quasi-periodic variability caused
by interaction of the circumbinary disk with the SMBH binary
\citep[e.g.][]{2008Natur.452..851V, 2015Natur.518...74G,
  2015MNRAS.453.1562G, 2015MNRAS.454L..21C}. Here we focus
on time-resolved UV spectroscopy.

Two spectroscopic techniques have been employed to identify
possible SMBH binaries. One class of methods measures the
``absolute'' shifts of the broad emission lines (BELs),
thought to be produced close to the black holes, relative to
the narrow emission lines, which are believed to have
redshifts close to that of the host galaxy
\citep[e.g.][]{2009Natur.458...53B,2010ApJ...720L..93D,
  2011ApJ...738...20T,2012ApJS..201...23E}. This method
tends to select a very rare sub-class of targets with large
velocity offsets in the BELs. Such an appearance may
well result from the orbital motion, but in many cases they
may be produced by alternative mechanisms \citep[e.g.][]
{2005PhDT.........5G, 2012ApJS..201...23E,
  2012ApJ...759...24S}. To resolve this ambiguity requires a
long baseline of observations.  In addition, many of these
objects exhibit variations in their BEL line shapes as the
luminosity varies \citep[e.g.][]{2013MNRAS.433.1492D}, which
may contaminate the velocity shift detections. {\bf A related
method uses binary BH models to predict kinematic signatures
of binarity in the broad lines \citep[see][]
{2016ApJ...828...68N}}.

The second type of search, utilized in this paper \citep[see
also][]{2013ApJ...777...44J, 2013ApJ...775...49S} uses
multiple observations of the same quasar to search for
time-varying velocity shifts that may indicate orbital
motion.  We focus here on $0.4 < z < 2$ quasars and use the
\ion{Mg}{II} $\lambda~2800~\ang$ line to trace the BEL
region.

\subsection{The Method}

Our basic binary BH search approach is to cross-correlate
multiple epochs of spectroscopic observations of the same
objects in search of radial velocity shifts that might
indicate binary orbital motion \citep[see e.g.][]
{1983LIACo..24..473G, 2016Ap&SS.361...59S}. We measure the
velocity shift $\Delta V$ between epochs with typical
time-baselines ranging from 1 to 14 years, and then attempt
to ascertain, using multiple epochs, whether $\Delta V$ can
be ascribed to orbital motion.  Thus, we are using the
broad-line region (BLR) of these quasars as the dynamical
tracer.  This method has both strengths and weaknesses that
set some fundamental limitations on what we can do.

The main strength is that we can target a large number of
accreting BHs at a cosmologically interesting time
($z \approx 2$), when merging activity peaked. Since there
are very few concrete observational limits on the lifetimes
of binary BHs, any tool that might yield interesting
limits must be explored.

The weakness is that we are depending on BLR physics that we
do not understand.  There are a number of uncertainties that
we must keep in mind as we evaluate the observational
results. First of all, we do not know the size of the BLR
precisely. At the point where the BLR is no longer
gravitationally bound to an individual BH, but rather
envelopes both, our method no longer works. Here we assume that
relationships between BLR size and AGN luminosity that have
been calibrated at lower redshift also apply to these
moderate redshift quasars, even if they harbor
binary BHs. Generally speaking, the BLR size is likely
$0.01-0.1~\pc$ for the quasars that we consider here
\citep{2010ApJ...725..249S}. Thus, we can only hope to probe
binaries with separations comparable to or larger than a
tenth of a pc.  At yet larger separations we do not have
long enough time baselines to be sensitive to orbital
velocity shifts. At smaller separations, the two BHs will
likely share a common BLR.

There is also the possibility that the BLR is completely
different in the case of binary BHs. There may be suppressed
accretion, different dynamics in the BLR due to tidal
truncation, or even no accretion at all on sub-pc scales. If
the BLR is different, then our BH scaling relations, which
already carry large systematic uncertainty
\citep[e.g.,][]{2013ApJ...775...49S}, may be even more
uncertain.  We therefore adopt a range of reasonable values
for $R_{\rm BLR}$ throughout the paper to reflect these
uncertainties.  If accretion is considerably lowered, then
our method has limited value. Still, given the lack of
observational constraints on SMBH binary evolution
timescales, it is worth searching for large radial velocity
shifts in the data.

The second uncertainty that we must contend with is that
BLRs may vary in velocity even in single AGN
\citep{2016arXiv160201975S}. In some dramatic cases, these
shifts are believed to arise from hot spots in the accretion
disk \citep[e.g.][]{1997ApJ...490..216E}. Even in typical
AGN, velocity shifts can arise as the AGN continuum varies,
depending on the illumination pattern of the BLR
\citep[][and discussion in \S 3.4]{2015ApJS..217...26B}. 
We do not know the full range
of velocities that may arise from this
reverberation. Therefore, these velocity shifts comprise a
significant source of noise for our method.  

In \citet{2013ApJ...777...44J}, we found that the
distribution in $\Delta \,V$ for two epochs, with $\sim 1$
yr time separations, had a width of $82~\km~\s^{-1}$, and
this width is very likely dominated by line variability in
AGN with single BHs that leads to measured velocity shifts
but is unrelated to orbital motion. Since we do not know the
full distribution of velocity shifts in single AGN as a
function of time, we must try to empirically determine this
distribution from the data we have at hand, which we will
discuss in \S\ref{sec:modeling} and \ref{sec:discussion}. It
is important to understand from the beginning that spectral
variability in the lines of single AGN constitutes a serious
source of contamination for us. On the other hand, as we
will show from our data, velocity shifts of
$\Delta V > 1000~\kms$ seem to be very rare from a single
AGN. Given the 10 year time baselines considered
here, we would be able to detect such dramatic velocity offsets
from radial velocity shifts in binary BHs. Thus, we
can draw some conclusions about the residence times of
binary BHs from these data despite contamination from single
AGN. 

\subsection{Structure of the Paper}

Our paper is slightly non-standard in that we present two
related (but distinct) experiments.  First, in
\S\ref{sec:ju-et-al} we present a third spectroscopic epoch
for 21 candidate binary BHs from
\citet{2013ApJ...777...44J}. Compared to this work we
increase the time baseline of monitoring by an order of
magnitude. This first experiment highlights very well how
velocity jitter in the BLRs of single BHs acts as a major
source of contamination for our method. In the short term,
we are forced to search for just the most extreme velocity
shifts that may result from long time-baseline monitoring.

Second, in \S\ref{sec:extension} we take advantage of
considerably more data from the SDSS III/BOSS survey, which
allows us to expand on the original work of
\citet{2013ApJ...777...44J} by adding a number of new
objects.  In particular, we are able to consider
$\sim 10~\yr$ time baselines for 10121 quasars in total
(1438 with high S/N ratio, see \S\ref{sec:extension}). 
{\bf With such long time baselines we can focus on the 
expected tail of $\Delta V > 1000~\kms$ objects 
expected from orbital motion.}

The remainder of the paper is structured as
follows. \S\ref{sec:method} explains the method we use for
data analysis. The results of our two experiments are
interpreted in \S\ref{sec:modeling}. \S\ref{sec:discussion}
discusses our findings and summarizes the paper.

\section{Cross-Correlation Analysis} 
\label{sec:method}

For both parts of our paper, our main method is to
cross-correlate multiple epochs of spectroscopy in search of
radial velocity shifts. Here we describe our main steps in
performing the cross-correlation.  We implement our analysis
routines in \verb|Python|, and make use of the least-squares
fitting package with the Levenberg-Marquardt algorithm,
\verb|lmfit|, wherever fitting is needed. To detect velocity
shifts, we focus on the temporal variations of broad
emission lines. We describe here the analysis we perform to
prepare the spectra for cross-calibration.

\subsection{Fitting and subtraction of the continuum
  emission}
\label{sec:fit-continuum}

Similar to \citet{2013ApJ...777...44J}, we first fit the
continuum emission and remove it. Continuum subtraction is
needed because either intrinsic variability in the continuum
shape or artifacts in the continuum between epochs can
contaminate the cross-correlation signal. We list the
significant emission lines that affect our continuum fitting
process in Table \ref{tab:emission-mask}.  Wavelength ranges
centered at $\lambda_c$ with half-width $\lambda_w$ are
masked. After this procedure, the continuum emission is
fitted by a 5th order polynomial, which empirically allows
for sufficient flexibility. This continuum-removal procedure
isolates the BEL from the quasar continuum.  We note that
all broad lines, including the \ion{Fe}{II} pseudocontinuum,
provide signal that in principle should trace the radial
velocity motions of the BLR, and thus we do {\it not} fit
and remove the broad Fe lines.

%%%%%%%%%%%%%%%%%%%%%%%%%%%%%%%%%%%%%%%%
\begin{deluxetable}{lcc}
  \tablecolumns{3} 
  \tabletypesize{\scriptsize}
  \tablewidth{0pt}
  \tablecaption{Emission line masks for continuum fitting.}
  \tablehead{
    \colhead{Species} & \colhead{Line center}
    & \colhead{Half-width} \\
    & \colhead{($\lambda_c/\ang$)}
    & \colhead{($\lambda_w/\ang$)}
    }
    \startdata
    \ion{O}{VI}		& 1035	& 35	\\
    Ly$\alpha$		& 1216	& 75	\\
    \ion{N}{V}		& 1241	& 35	\\
    \ion{Si}{IV}	& 1398	& 35	\\
    \ion{C}{IV}		& 1549	& 35	\\
    \ion{He}{II}	& 1640	& 35	\\
    \ion{C}{III}	& 1909	& 35	\\
    \ion{Mg}{II}	& 2799	& 80	\\
    \ion{O}{II}		& 3727	& 35	\\
    H$\delta$		& 4102	& 80	\\
    H$\gamma$		& 4340	& 80	\\
    \ion{O}{III}	& 4363	& 35	\\
    $\hb$   		& 4861	& 100	\\
    \ion{O}{III}	& 4959	& 35	\\
    \ion{O}{III}	& 5007	& 35	\\
    \ion{O}{I}		& 6350	& 35	\\
    \ion{O}{I}		& 6364	& 35	\\
    \ion{N}{II}		& 6548	& 35	\\
    H$\alpha$		& 6563	& 330	\\
    \ion{N}{II}		& 6583	& 35	\\
    \ion{S}{II}		& 6716	& 35	\\
    \ion{S}{II}		& 6734	& 35
    \enddata
  \label{tab:emission-mask}
\end{deluxetable}
%%%%%%%%%%%%%%%%%%%%%%%%%%%%%%%%%%%%%%%%

After removal of the continuum, we extract the \ion{Mg}{II}
BEL feature from the spectrum in the rest-frame wavelength
window ($2698~\ang < \lambda < 2798~\ang$ or
$2748~\ang < \lambda < 2848~\ang$). The spectrum of each
object is converted into the rest frame using the redshift
value of the first epoch observation, since there is
significant variance between the redshift measurements for
different epochs: the mean and standard deviation of
$(z_\mathrm{DR7}-z_\mathrm{DR12})$ are 0.00032 and 0.068,
respectively. Discrepant redshift data are indications of
the quasars' intrinsic variations. {\bf By using the redshift of
the first epochs uniformly, we are able to measure any radial velocity 
shifts without additional noise from the redshift measurements.}

\subsection{Cross-correlation identification of velocity
  shifts}
\label{sec:cross-corr}

\begin{figure*}
  \centering
  \includegraphics[width=6.3in, keepaspectratio]
  {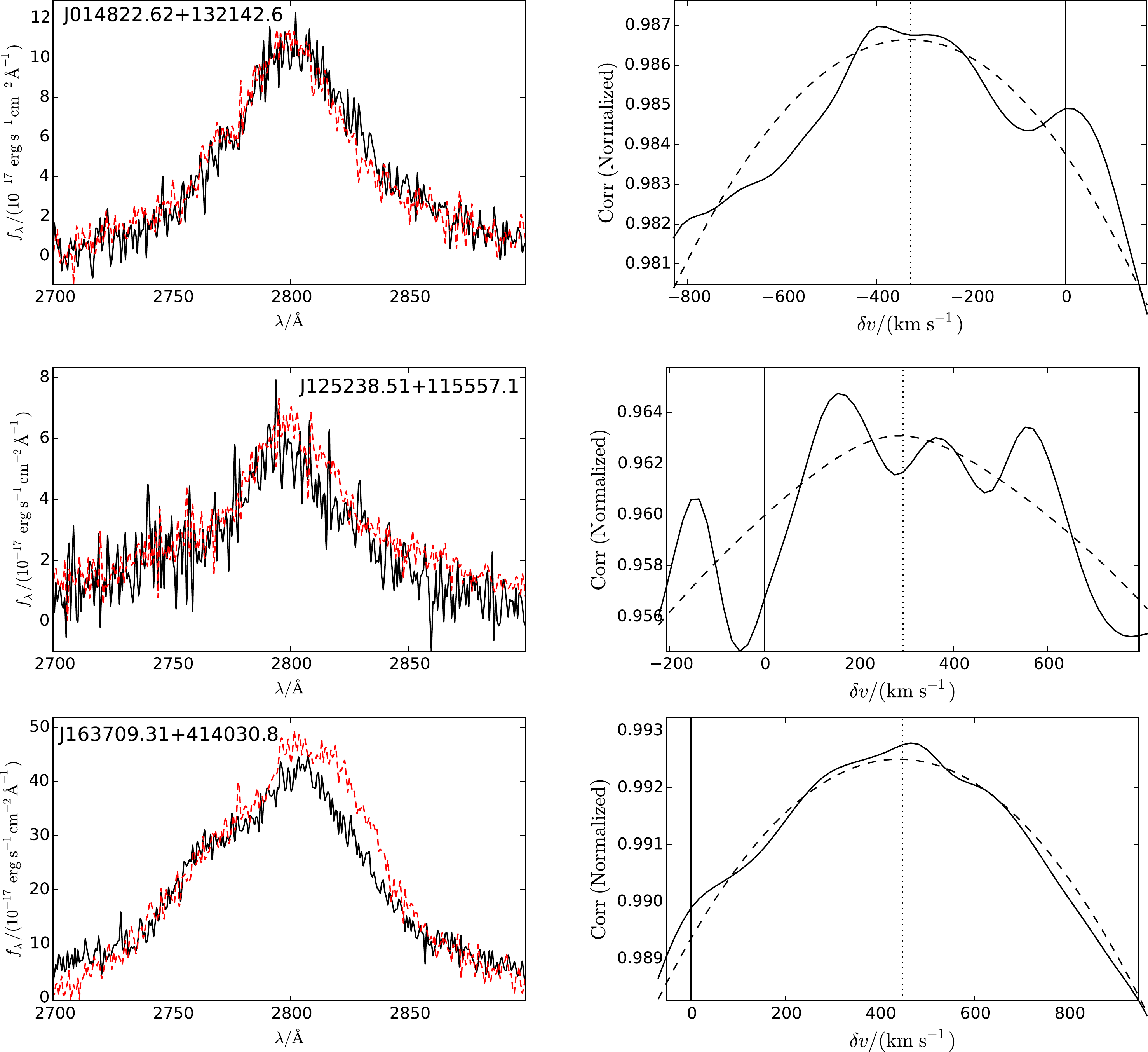}
  \caption{
 % {\bf these are not just examples, 
 % they are the ones with %the largest shifts. best to 
 % explain this here.}
     Examples of the cross-correlation between two epochs 
     of spectra of the \ion{Mg}{II} line, showing three
    of the candidates with the most prominent line shifts (see Table
    \ref{tab:high-v-high-snr}): J014822.62+132142.6 (top
    row), J125238.51+115557.1 (middle row), and
    J163709.31+414030.8 (bottom row). 
    {\bf Left column}: comparison of spectra. The black lines
    represent the observation of the first epoch, and the
    red dashed lines represent the second.  {\bf Right
      column}: the cross-correlation function (see equation
    \ref{eq:cross-corr}). The solid curve shows results
    directly measured from the spectra, while the dashed
    curve is the result using Gaussian fitting. The maximum
    of the fitted curve is indicated by a vertical dotted
    line, and $\delta v = 0$ is indicated by a vertical
    solid line.}
  \label{fig:spec_example}
\end{figure*}

Taking the continuum-subtracted spectra as input, we now
evaluate the cross-correlation function between the spectra
with the maximum time separation. We evaluate the normalized
cross-correlation function, using the following formula
\citep[e.g.][]{2013ApJ...777...44J},

\begin{equation}
  \label{eq:cross-corr}
  \begin{split}
    \mathrm{Corr}(D) & = \left[\sum_{i=0}^N ( a_i -
      \bar{a})^2 \right]^{-1/2} \left[ \sum_{i=0}^N ( b_i -
      \bar{b})^2
    \right]^{-1/2} \\
    & \times \left[ \sum_{i=0}^{N-|D|+1} (a_{i+D} - \bar{a})
      ( b_i - \bar{b}) \right] \ ,
  \end{split}
\end{equation}
where $D$ is the number of ``delayed'' channels, and $a_i$
and $b_i$ are the measured spectral fluxes from the two
epochs of observation of the $i$th channel.  The SDSS and
BOSS spectrographs are different \citep[see][]
{2013AJ....146...32S}, and we interpolated the DR12 spectra
onto the channel grid of DR7 spectra to conduct sensible
correlation calculations. The value of $\mathrm{Corr}(D)$
peaks at $D=0$ when the two spectra are identical.
Results of these cross-correlation procedures are
illustrated by, as an example, Figure \ref{fig:spec_example}.

Before feeding the spectra into equation
\eqref{eq:cross-corr}, 
we ``oversample'' the original spectrum by a factor of
8. This procedure does not introduce any extra information,
but it makes the correlation function smoother and eases the
peak-finding procedure. The velocity shift represented by
each oversampled channel (i.e. the resolution of our
cross-correlation scheme) is
$\delta v_\mathrm{res} = 8.6~\km~\s^{-1}$.

In general, because the \ion{Mg}{II} lines are smooth and
broad, we expect the cross-correlation function to be
dominated by a single peak. Exceptions are produced by broad
absorption lines (BALs; \citealt{2007ApJ...656...73L,
  2008ApJ...675..985G, 2010ApJ...713..220G}) but we have
removed these features by hand.
% (in 52 objects with $\snr > 3$ and $|\Delta V| > 
% 200~\km~\s^{-1}$, we picked out 16). 
Other small bumps may be caused by narrow absorption lines 
or noise features, but they do not impact our ability to 
identify the primary cross-correlation peak.

We aim to find the maximum in the {\it overall} structure of
the correlation function. This is different from
\citet{2013ApJ...777...44J}, in which the authors
used a somewhat smaller window to determine the cross correlation 
maximum. By trial and error we found that using the
sum of three models to fit the maximum works best. We fit
the cross-correlation function with a combination of two
skewed Gaussian models (with positive-definite amplitude,
providing skewness to our fitting function), and a
second-order polynomial to the residual continuum. 
It is straightforward to identify the maximum of the cross-correlation 
signal using the fit. 
The location of the fit maximum, $\Delta\lambda$, is
directly used to calculate the value of the velocity shift,
\begin{equation}
  \label{eq:v-shift-corr}
  \Delta V \simeq \dfrac{\Delta\lambda}
  {\lambda_{\ion{Mg}{II}}} c\ ,
\end{equation}
where $c$ is the speed of light.

\section{Experiment 1: Ju et al. Follow-up}
\label{sec:ju-et-al}

\subsection{Sample properties}
\label{sec:sample-properties}

\citet{2013ApJ...777...44J} analyzed 4024 QSOs with two
epochs of observation in the quasar catalog of the seventh
data release (DR7) of Sloan Digital Sky Survey
\citep[SDSS][]{2000AJ....120.1579Y,
  2010AJ....139.2360S}. For completeness, here we briefly
summarize their final sample selection, since this paper
analyzes additional observations of their targets.

In order to focus on the \ion{Mg}{II} emission line, objects
were selected in the redshift range $0.36 < z < 2.0$, with
$g$-band absolute PSF magnitudes
ranging from $-18.4$ to $-27.5$ mag (median $-24.1$ mag).
The typical time lag between observations ranged from
$0.003~\yr$ to $7.03~\yr$, with mean value $0.7~\yr$.  Based
on broad emission line scalings \citep{2011ApJS..194...45S},
the virial BH masses range from $10^{7}~M_\odot$ through
$10^{10.5}~M_\odot$ (peaked at $10^{8.65}~M_\odot$).
\citet{2013ApJ...777...44J} focused on the 1523 objects with
high signal-to-noise (S/N) ratios ($\snr > 10$; each pixel 
corresponds to $70~\kms$).

Using a cross-correlation between the two epochs (see
\S\ref{sec:cross-corr} below), the distribution of
velocity shifts was measured first for all objects,
regardless of time separation, and found to have a
dispersion of $\sigma = 101~\km~\s^{-1}$ for the whole
sample, and $\sigma = 82~\km~\s^{-1}$ for the high $\snr$
subsample.
% We find a broad distribution of velocities with a
% FWHM of $80~\kms$.
Just requiring velocity shifts greater than $>3 \sigma$
yielded 7 high-confidence candidates, and 64 extra possible
candidates.

Twenty-one (21) of these candidates were revisited as part
of the TDSS (Time-Domain Spectroscopic Survey, a part of
SDSS III), which provides spectroscopic data for
time-variable objects, including roughly 135,000 quasars
\citep[see][]{2015ApJ...806..244M,2015ApJS..221...27M,
  2016arXiv160202752R}.  With this additional epoch, we now
have much longer observed time baselines for these objects,
ranging from $12.91$ through $14.31~\yr$ (median
$14.16~\yr$), compared to the original $0.17~\yr$ to
$6.16$~yr (median $0.98~\yr$). Furthermore, we now have a
total of three epochs. These observations not only
yield more reliable velocity shift detections by probing
longer timescales, but also trace the evolution of these
systems in order to address whether the velocity shifts that
were detected in \citet{2013ApJ...777...44J} are consistent
with the binary BH picture.

\subsection{Verification criteria}
\label{sec:verify-criteria}

The data analysis algorithm described in \S \ref{sec:method}
is different from the method presented in
\citet{2013ApJ...777...44J}, although the two are closely
related. To be sure that we can analyze all three
epochs consistently, we first apply our analysis routines
to the two-epoch spectra that were examined in
\citet{2013ApJ...777...44J} to verify that the velocity
shifts are detected with similar values. Focusing on the
selected candidates specified in
\citet{2013ApJ...777...44J}, we analyzed the spectral data
acquired during the first two epochs using the procedures
described in \S\ref{sec:fit-continuum} and
\S\ref{sec:cross-corr}, and reproduced the
\citet{2013ApJ...777...44J} $|\Delta V|$ measurements
consistently. As a sanity check, we re-apply our scheme
to the candidates in \citet{2013ApJ...777...44J} and find
that they all remain candidate binaries under
this new algorithm.

As an additional sanity check, in addition to \ion{Mg}{II},
we also examine velocity shifts measured with the $\hb$ BEL
if applicable. We exclude those with $\snr<1$ in the
wavelength window surrounding $\hb$ BEL (see
\S\ref{sec:extension} for the definition of $\snr$). $\hb$
does come with additional complications due to the blending
of [\ion{O}{III}]$\lambda\lambda~4959, \, 5007$. Values of
$\Delta V$ measured by $\hb$ and \ion{Mg}{II} do exhibit
some level of consistency, but the $\hb$ results still
suffer from considerable uncertainties due to blending with
narrow lines, causing a good deal of scatter between the two estimates.

\subsection{The additional epoch}

After analyzing the first two epochs consistently, we now
turn to the third epoch. For the follow-up observations, we
use a straightforward criterion to determine whether the
original velocity shift can be ascribed to orbital
motion. For a secondary SMBH in a binary system, we define
the empirical acceleration along the line-of-sight (LoS)
$a_\los$,
\begin{equation}
  \label{eq:alos-def}
  a_\los \equiv \dfrac{\Delta V}{\Delta t}
  = \dfrac{\Delta V}{(1+z)\Delta \tau}\ ,
\end{equation}
where $\Delta t$ is the time baseline in the observer's rest
frame, and $\Delta \tau$ is its counterpart in the rest
frame of the object. In what follows, we will use $t$ for
the time variable in the observer's frame and $\tau$ for
time in object's frame. This ``acceleration'' magnitude
should not be thought of as a physical quantity, since
$\Delta V$ is measured in the rest frame of the object, but
simply as a convenient way to test the consistency of the
different epochs.

The typical time-scale of variation is evaluated, regarding
the time variability of $a_\los$ and $\Delta V$. We assume
that the secondary (the smaller BH mass in the
binary system) carries the BLR, and that the virial mass
inferred from the BEL width indicates the mass of the
secondary. Hence, in what follows we use $M_\bh$ to denote
the mass of secondary.  We also assume that the secondary
travels along a circular orbit by $\pi/2$ phase (which
covers a full amplitude of LoS motion). Considering time
dilation, the time it takes the binary to cover a quarter of
its orbit is:
\begin{equation}
  \label{eq:t_a_var}
  \begin{split}
    \Delta t_a & = \dfrac{\pi(1+z)}{2\Omega} \\
    & = 23.4~\yr \times 
    (1+z) \left( \dfrac{M_\bh/q}{10^9M_\odot} \right)^{-1/2}
    \left( \dfrac{r}{0.1~\pc} \right)^{3/2}\ ,
  \end{split}
\end{equation}
where $\Omega$ is the orbital angular speed assuming a
circular orbit,
\begin{equation}
  \label{eq:omega_orbit}
  \Omega \equiv \left| \dfrac{G M_\bh (1 + q)}
    {r^3 q} \right|^{1/2}\ ,
\end{equation}
and $q$ is the mass ratio of the secondary to primary BH.

Our velocity shift detection time baseline is around
$\Delta t \, \sim 10~\yr$. The separation $r$ between the
two BHs must be $\sim 0.1~\pc$ or smaller in order to detect
the velocity shifts in $\sim 1$ yr, as set by
\citet{2013ApJ...777...44J}. In that case, we
would expect $\Delta t_a$ to be comparable to $\Delta t$.
We would also expect $a_\los$ to maintain a similar
amplitude when we inspect the ensemble of binary system as a
whole. Thus, a relatively steady value of $a_\los$ is the
criterion that we adopt to determine whether the velocity
shifts observed in \citet{2013ApJ...777...44J} can be
attributed to binary orbital motion.

On the other hand, it is worth noting that orbital motion
could significantly change the value and even potentially
the direction of $a_\los$ for a BH binary with a high mass
and a small orbital separation. Also, there may be
cases where the velocity shift is significant during a short
epoch, but small in the long run. Indeed, consider a
BH binary on a circular orbit with initial orbital phase $\phi$,
angular speed $\Omega$, orbital velocity $V_\mathrm{co}$,
and inclination angle $i$. After a period of time
$\Delta t$, the magnitude of velocity shift is
\begin{equation}
  \label{eq:dv-timescale-change}
  % \begin{split}
  \dfrac{|\Delta V|}{\Delta \tau}
  = \dfrac{2 V_\mathrm{co}}{\Delta \tau}
  \left| \sin \left(\phi + \dfrac{\Omega\Delta \tau}{2}
    \right) \sin \left(\dfrac{\Omega\Delta \tau}{2}\right)
    \sin i \right| \ .
  % \end{split}
\end{equation}
Given a special initial phase $\phi$ and inclination angle
$i$, the apparent acceleration can be as large as
$\sim\Omega V_\mathrm{co}$ in the short term. As $\Delta t$
increases, the value of acceleration will drop
significantly, being no greater than
$2 V_\mathrm{co}/\Delta \tau$. However, we note that this
will only happen under specific initial conditions, and
cannot dominate the sample statistically.

\subsection{Results}
\label{sec:follow-up}

We measure the velocity shifts of the candidates from DR7,
presented by \citet{2013ApJ...777...44J}. Table
\ref{tab:follow-up} compares and updates the results for the
\citet{2013ApJ...777...44J} objects using the SDSS IV
spectra acquired for this purpose
\citep{2015arXiv150804473D}. It is immediately clear that
the velocity shifts are not consistent with our expectations
for orbital motion. The inferred $a_\los$ values are lower
than expected, often by an order of magnitude. The magnitude
of the apparent acceleration is always decreasing over the
longer time baseline. If the velocity shifts were due to
orbital motion, one would expect to observe that a fraction
of $a_\los$ measurements increase over time, which is not
observed here.

\begin{figure}
  \centering
  \includegraphics[width=3.1in, keepaspectratio]
  {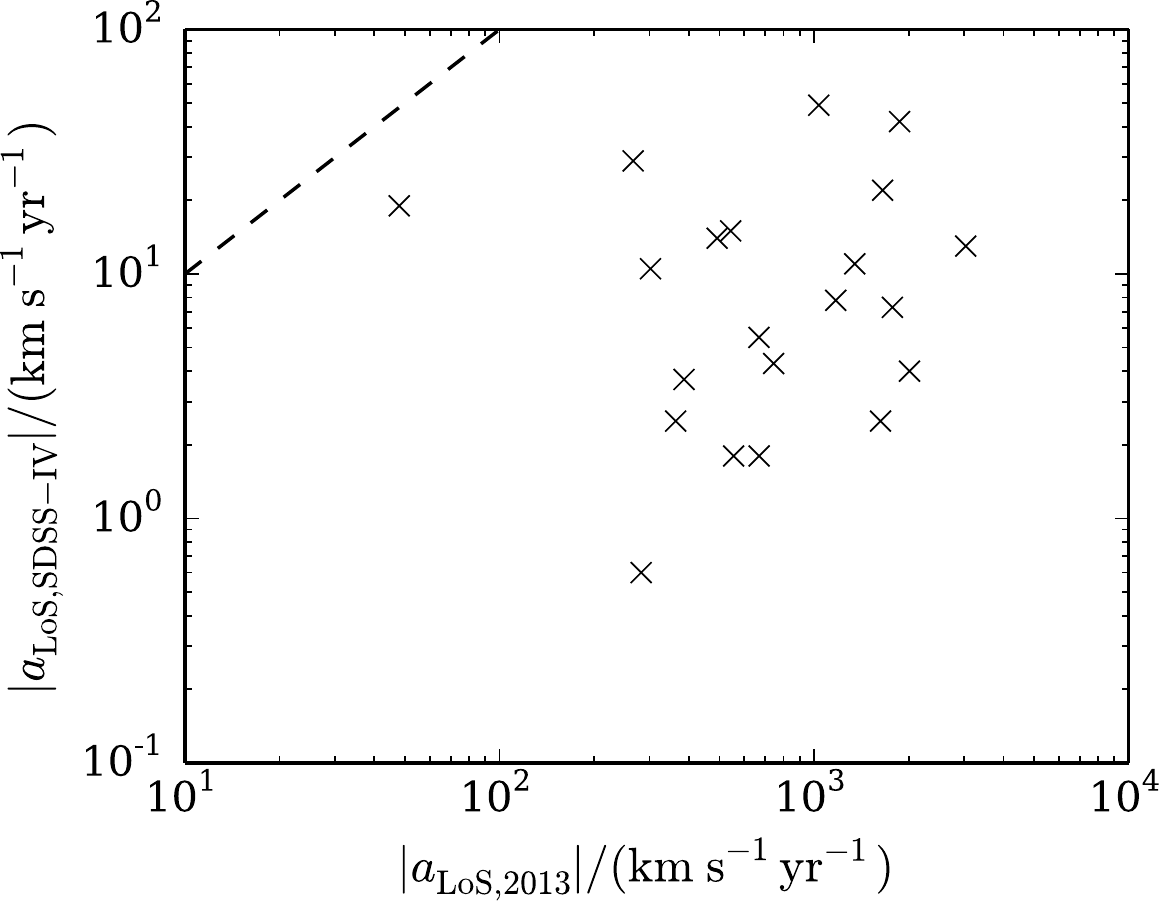}
  \caption{Comparison of the LoS acceleration measurements.
    Each data point indicates a row in Table
    \ref{tab:follow-up}. The horizontal axis indicates
    $a_\los$ measured in \citet{2013ApJ...777...44J} (median
    time baseline $0.98~\yr$) , and the vertical axis
    indicates $a_\los$ obtained from the pertinent SDSS-IV
    data (median time baseline $8.04~\yr$). The dashed line
    on the upper-left corner is the line of equal value of
    the two $a_\los$.  }
  \label{fig:acc_compare}
\end{figure}

Instead, we find comparable values of $\Delta V$ between the
first and second epoch (typically separated by one year) and
between the first and third epoch (typically separated by
eight years). The correlation coefficients between the two
different $|\Delta V|$ measurements are also poor: 0.47 for
$a_\los$, and 0.26 for $\Delta V$.  These results reveal
that the detections are highly unlikely to be attributed to
binary orbital motion (see \S\ref{sec:verify-criteria}). A
comparison of $|a_\los|$ values is presented in Figure
\ref{fig:acc_compare}.

We consider various explanations for the large velocity offsets, that 
are clearly not caused by orbital motion. One possibility is 
just noise spikes or imperfect sky subtraction.
In \citet{2013ApJ...777...44J}, we tried to reproduce the
distribution of acceleration values using simulations of the
random (e.g., photon noise) and systematic (e.g., sky
subtraction) noise in the spectra. From these factors alone,
we were unable to reproduce such a broad distribution. Thus,
we conclude that the velocity offsets we measure are real,
but are not due to binary motion.  We infer that the
velocity shifts should be attributed to real spectral
variability, but arising from single AGN.

In the literature, velocity changes in the BLR have long
been seen \citep[e.g.][]{1997ApJ...490..216E,
  2003ApJ...598..956S, 2004A&A...422..925S,
  2007ApJS..169..167G, 2010A&A...517A..42S,
  2010ApJ...718..168J, 2013MNRAS.433.1492D,
  2014AJ....147...12B, 2014A&A...572A..66P,
  2016ApJ...817...42L}. In some cases these are interpreted
in the context of accretion disk models
\citep[e.g.][]{2008ApJS..174..455B}. In others the velocity
shifts can be attributed just to reverberation
\citep{2014AJ....147...12B,2015ApJS..217...26B}.  As the
continuum luminosity of the AGN varies with time, so too
does the emission from photoionized gas, with a time
lag. Asymmetries in the line transfer function between the
continuum and line emission light curves will cause apparent
velocity shifts in the BLR. \citet{2015ApJS..217...26B} show
that these shifts can extend to at least a couple hundred
$\km~\s^{-1}$. Thus far, only relatively low-luminosity AGN
have been monitored with high enough cadence to yield a
constraint here, and in these cases the empirical velocity
shifts never go higher than a few hundred $\kms$. We do not
have a model for the the spectral variability due to
reverberation or other causes \citep[e.g.][]
{2013ApJ...763L..36L, 2016Ap&SS.361...59S}, but our
simulations suggest that these physical effects dominate the
velocity shifts that we measure.

\subsection{Summary}

From our follow-up of the \citet{2013ApJ...777...44J}
candidates, we conclude that the broad-line variability in
single AGN is a significant source of contamination in our
search for binary candidates. Unfortunately, we do not yet
know the full distribution of line shifts that is expected
from reverberation. Nor do we know whether the distribution
of velocities is a function of other properties of the AGN,
such as BH mass or luminosity. We are not in a position to
model and remove the contribution to the measured velocity
shifts from these reverberation and disk variability
effects. However, velocity shifts $>500~\kms$ have not been
seen in single objects. Given the 10-yr time baselines that
we will present in the next section, we expect to see
$>1000~\kms$ lineshifts from orbital motion, which we do not
believe can arise from single BHs. However, in the longer
term, we must measure the distribution of line shifts from
reverberation (e.g., using ongoing multi-object
reverberation mapping campaigns;
\citealt{2014Natur.513..210S,2015ApJS..216....4S}) so that
we can more robustly isolate the binary candidates.

% Notes on papers: Simic & Popovic 2016: simple models,
% included above Double-peaked things: Gaskell 1984,
% Popovic+2000,Eracleous+2012,Bon+2012, Bogdanovic 2015,
% Runnoe+2015 Eracleous+1997, Decarli+2010a,Shapovalova+
% 2015

% From extensive testing, we do not believe that
% straightforward noise issues can explain these shifts
% \citep[see discussion in][]{2013ApJ...777...44J}. We will
% discuss possible causes for the velocity jitter in more
% detail in \S\ref{sec:discussion}.

%%%%%%%%%%%%%%%%%%%%%%%%%%%%%%%%%%%%%%%%
\begin{deluxetable*}{lccccccccc}
\tablecolumns{10} 
\tabletypesize{\scriptsize}
\tablewidth{0pt}
\tablecaption{Follow-up measurement of velocity shift}
\tablehead{ 
  \colhead{SDSS ID} &
  \colhead{$z$} &
  \colhead{$\Delta t_0$} &
  \colhead{$\Delta V_{0}$} &
  \colhead{$a_{\los,0}$} &
  \colhead{$\Delta t_1$} &
  \colhead{$\Delta V_{1,\ion{Mg}{II}}$} &
  \colhead{$a_{\los,1,\ion{Mg}{II}}$} &
  \colhead{$\Delta V_{1,\hb}$} &
  \colhead{$a_{\los,1,\hb}$} 
  \\
  \colhead{} &
  \colhead{} &
  \colhead{($\yr$)} &
  \colhead{($\km~\s^{-1}$)} &
  \colhead{($\km~\s^{-1}~\yr^{-1}$)} &
  \colhead{($\yr$)} &
  \colhead{($\km~\s^{-1}$)} &
  \colhead{($\km~\s^{-1}~\yr^{-1}$)} &
  \colhead{($\km~\s^{-1}$)} &
  \colhead{($\km~\s^{-1}~\yr^{-1}$)}
  \\
  \colhead{(1)} & \colhead{(2)} & \colhead{(3)} & 
  \colhead{(4)} & \colhead{(5)} & \colhead{(6)} & 
  \colhead{(7)} & \colhead{(8)} & \colhead{(9)} &
  \colhead{(10)}
}
\startdata
J002028.34+002915.0  & 1.93 & 0.2301 & 310 & 1347
& 14.15 & $-$155.5 & $-$11.0 &  &  \\
J002311.06+003517.5  & 0.42 & 0.23 & 698 & 3035
& 14.15 & 181.0 & 12.8 &  &  \\
J002411.66+004348.1  & 1.79 & 1.15 & 444 & 386
& 14.17 & $-$51.8 & $-$3.7 &  & \\
J002444.12+003221.2  & 0.40 & 0.23 & 380 & 1652
& 14.21 & 309.6 & 21.8 & 154.8 & 10.9 \\
J003451.86$-$011125.6  & 1.84 & 1.28 & $-$362 & $-$283
& 14.28 & $-$8.6 & $-$0.6  &  &  \\
J004052.15+000057.3  & 0.41 & 1.28 & $-$1501 & $-$1173
& 14.31 & 112.1 & 7.8 & 293.1 & 20.5 \\
J004918.98+002609.5  & 1.94 & 0.28 & $-$290 & $-$1036
& 14.26 & $-$698.2 & $-$49.0 & &  \\
J011310.39+003113.3  & 0.41 & 1.12 & 552 & 493
& 14.27 & 198.3 & 13.9 & 190.1 & 13.3 \\
J013418.19+001536.7  & 0.40 & 1.05 & $-$1863 & $-$1774
& 14.22 & 103.7 & 7.3 & 0.0 & 0.0  \\
J014209.72+002348.3  & 1.35 & 1.14 & $-$414 & $-$363
& 14.01 & $-$34.6 & $-$2.5 &  &  \\
J014415.13+002349.7  & 1.71 & 1.14 & 345 & 303
& 14.01 & 146.5 & 10.5 & &  \\
J014905.16+005925.4  & 1.00 & 1.12 & $-$299 & $-$267
& 14.27 & $-$414.7 & $-$29.1 &  &  \\
J015454.88+004043.8  & 1.65 & 0.9 & 670 & 744
& 14.06 & 60.3 & 4.3 &  &  \\
J020527.77+005747.6  & 1.24 & 0.18 & 293 & 1628
& 14.10 & 34.6 & 2.5 &  &  \\
J020646.97+001800.6  & 1.68 & 0.18 & 362 & 2011
& 14.18 & $-$51.8 & $-$3.7 & &  \\
J025257.17$-$010220.7  & 1.25 & 0.17 & 318 & 1871
& 14.16 & 594.8 & 42.0 &  &  \\
J025316.47+010759.8  & 1.03 & 0.98 & $-$533 & $-$544
& 14.16 & $-$215.5 & $-$15.2 &  &  \\
J025331.93+001624.8  & 1.83 & 0.98 & 655 & 668
& 14.16 & $-$77.6 & $-$5.5 & & \\
J082012.63+431358.4  & 1.07 & 0.68 & 378 & 556
& 13.98 & 25.9 & 1.8 &  &  \\
J082214.83+431701.9  & 0.97 & 0.68 & $-$455 & $-$669
& 13.98 & 25.9 & 1.8 &  &  \\
J095656.44+535023.3  & 0.61 & 6.16 & 300 & 49
& 12.91 & 241.9 & 18.7 & $-$86.2 & $-$6.7
\enddata
\tablecomments{
  (1) SDSS ID of the object ($\alpha_{2000}$ and
  $\delta_{2000}$). 
  (2) Spectroscopic redshift of the object.
  (3) Time baseline in \citet{2013ApJ...777...44J}.
  (4) The velocity shift measured by
  \citet{2013ApJ...777...44J} with cross-correlation of
  rest frame wavelength window near the \ion{Mg}{II} BEL
  (see subsection \ref{sec:cross-corr}).
  (5) The LoS component of acceleration based on
  columns (3) and (4).
  (6) Time baseline of the SDSS IV data.
  (7) Velocity shift measured with cross-correlation of
  rest frame wavelength window near the \ion{Mg}{II} BEL.
  (8) The LoS acceleration of SDSS IV follow-up \ion{Mg}{II}
  measurement.
  (9) Velocity shift measured with SDSS IV data in the
  rest frame wavelength window $4761~\ang < \lambda < 4961\
  \ang$ near the $\hb$ BEL (if available).
  (10) The LoS acceleration based on $\hb$ BEL measured with
  SDSS IV data.}
\label{tab:follow-up}
\end{deluxetable*}
%%%%%%%%%%%%%%%%%%%%%%%%%%%%%%%%%%%%%%%% 

\section{Experiment 2: BOSS Quasars}
\label{sec:extension}
\subsection{Sample properties}

In addition to revisiting the \citet{2013ApJ...777...44J}
sample (\S\ref{sec:ju-et-al}), we also build on their sample
by selecting objects where the second epoch of observations
was carried out by the Baryon Oscillation Spectroscopic
Survey, acquiring spectroscopic redshifts of luminous red 
galaxies and quasars to measure cosmic large scale
structure, and hence baryonic acoustic oscillations
\citep[][]{2006AJ....131.2332G, 2011AJ....142...72E,
  2013AJ....145...10D, 2012ApJS..199....3R,
  2013AJ....146...32S}.  In the SDSS DR12 quasar catalog
\citep{2014A&A...563A..54P, 2015ParisDR12QSO}, the flag
``\verb|SDSS_DR7|'' is assigned to each object to indicate
whether this object has already been observed in DR7. We
specifically note that, unlike many of the objects in
\citet{2013ApJ...777...44J}, none of the objects analyzed in
this paper are re-observed because of observational problems
(i.e., those flagged as ``bad'' or ``marginal'').

% We confirm the correctness of this flag by spatially
% searching every DR7 QSO in the DR12 catalog, using the
% $k$-d tree method implemented by the \verb|KDTree| module
% in \verb|scipy| library. Among 25275 objects that have
% positive ``\verb|SDSS_DR7|'' flags, 25270 survive this
% cross-check.

We set the redshift range by requiring that there be at
least a $\pm 50~\ang$ window around the
\ion{Mg}{II}$~\lambda 2800~\ang$ line in the
$3800~\ang < \lambda < 9200~\ang$ wavelength range of the
SDSS spectra, leading to a range of $0.38 < z < 2.23$. A
total of 10121 quasars satisfy this redshift condition. The
minimum window of $\pm 50~\ang$ allows us to measure
velocity shifts up to $\sim 5353~\km~\s^{-1}$; this value
is far beyond what we actually detect, as we will see in
section \ref{sec:dr12-observation}.

\begin{figure*}
  \centering
  \includegraphics[width=7.1in, keepaspectratio]
  {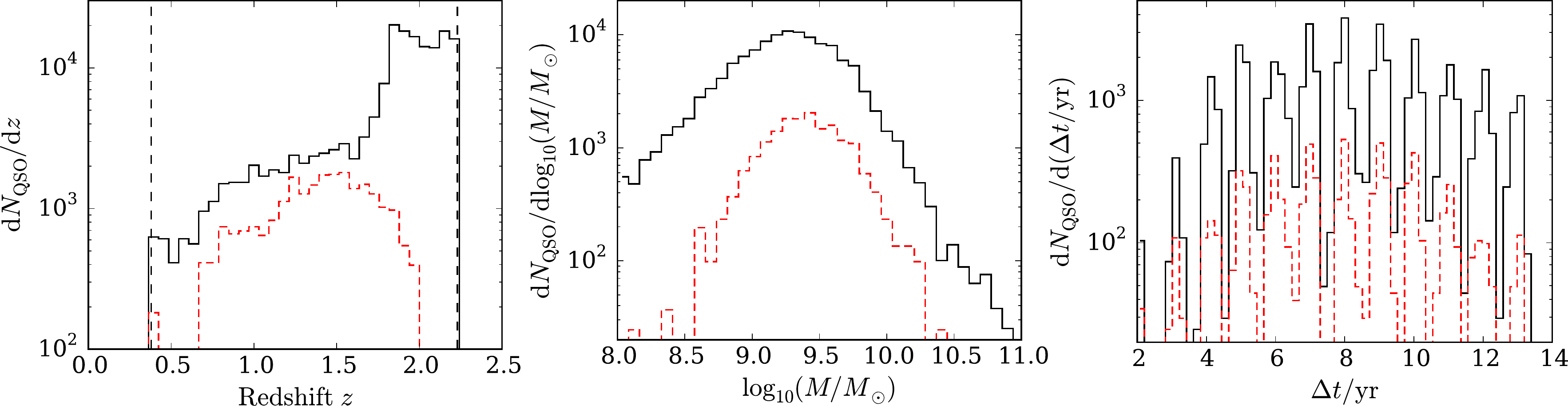}
  \caption{Basic properties of our SDSS DR7-DR12
    multi-observation sample. In each panel, the black
    histogram shows the distribution of our whole sample,
    and the red dashed histogram displays that of the
    $\snr > 3$ subsample. We present the statistics only for
    $0.38 < z < 2.23$ objects, which have their \ion{Mg}{II}
    BEL in the SDSS spectra.  {\bf Left panel}: Redshift
    ($z$); vertical dashed lines indicate the edges of
    \ion{Mg}{II} redshift window.  {\bf Middle panel}:
    Logarithm of BH mass based on
    \citet{2011ApJS..194...45S} modeling of FWHM of
    \ion{Mg}{II} BEL.  {\bf Right panel}: Observed time
    baseline of our velocity shift detections.  }
  \label{fig:boss_sample_property}
\end{figure*}

In Figure \ref{fig:boss_sample_property}, we present some
basic properties of the BOSS multi-epoch sample and the
high-S/N subsample ($\snr > 3$). The value of $\snr$ is
defined in a different way here as compared to
\citet{2013ApJ...777...44J}. In this work, we evaluate the
S/N in the rest-frame wavelength window surrounding the
\ion{Mg}{II} BEL ($2698~\ang < \lambda < 2798~\ang$
typically, but at least $2748~\ang < \lambda < 2848~\ang$)
{\it after} subtracting the continuum (see \S
\ref{sec:fit-continuum} for continuum subtraction). There
are 1438 high-S/N objects, out of the 10121 QSOs that fall
in the redshift window.

We span a range of observed time baselines from $1.91~\yr$
to $13.34~\yr$ with a median of $8~\yr$, considerably longer
than the mean time difference of $0.7~\yr$ observed by
\citet{2013ApJ...777...44J}. The absolute magnitude in the
$i$-band \citep{1996AJ....111.1748F} ranges from $-21.11$
through $-29.79$ mag with median $-26.56$ mag.  Using
standard scalings between broad-line width, AGN luminosity,
and BH mass \citep{2011ApJS..194...45S}, we find a median BH
mass of $1.8 \times 10^9~M_\odot$, with a standard deviation
of $1.24$ dex in log BH mass. Because of ambiguity in the
BLR dynamics in binary systems, we will adopt a range of BH
mass values (see section \ref{sec:modeling}). The median
redshift is $\langle z \rangle = 1.92$, with standard
deviation 0.39 (Figure \ref{fig:boss_sample_property}).

\subsection{Results}
\label{sec:dr12-observation}

\begin{deluxetable*}{lccccc}
\tablecolumns{7} 
\tabletypesize{\scriptsize}
\tablewidth{0pt}
\tablecaption{High-$\Delta V$ and high-SNR subsample based
  on BOSS data}
\tablehead{
  \colhead{SDSS ID} &
  \colhead{$z$} &
  \colhead{$\log_{10}M_\bh$} &
  \colhead{$\Delta t$} &
  \colhead{$\Delta V$} &
  \colhead{$a_\los$}
  \\
  \colhead{} &
  \colhead{} &
  \colhead{($M_\odot$)} &
  \colhead{($\yr$)} &
  \colhead{($\km~\s^{-1}$)} &
  \colhead{($\km~\s^{-1}~\yr^{-1}$)}
  \\
  \colhead{(1)} & \colhead{(2)} & \colhead{(3)} & 
  \colhead{(4)} & \colhead{(5)} & \colhead{(6)}
}
\startdata
J002127.88+010420.2 & 1.82 & 9.81 & 9.81 & $-$345 & $-$35 \\
J003333.61$-$001858.1 & 0.69 & 9.27 & 10.03 & 293 & 29 \\
J014822.62+132142.6 & 0.88 & 9.36 & 10.33 & $-$345 & $-$33 \\
J091344.40+150935.1 & 0.94 & 9.45 & 5.28 & $-$276 & $-$52 \\
J095929.88+633359.8 & 1.85 & 10.74 & 11.14 & $-$328 & $-$29 \\
J103623.66+152733.3 & 1.92 & 9.95 & 5.02 & 328 & 65 \\
J105611.27+170827.5 & 1.33 & 9.62 & 5.93 & $-$293 & $-$49 \\
J110038.79+282036.1 & 1.79 & 9.60 & 7.06 & 258 & 36 \\
J125238.51+115557.1 & 1.85 & 9.64 & 6.87 & 310 & 45 \\
J133615.79+495529.0 & 1.50 & 9.62 & 8.13 & $-$207 & $-$25 \\
J135109.38+320049.0 & 1.12 & 9.64 & 4.85 & $-$328 & $-$67 \\
J135218.49+224708.9 & 1.45 & 10.09 & 4.24 & $-$397 & $-$93 \\
J163600.37+432802.7 & 0.94 & 9.42 & 11.00 & 276 & 25 \\
J163709.31+414030.8 & 0.76 & 9.76 & 10.99 & 448 & 40
\enddata
\tablecomments{
  (1) SDSS ID of the object.
  (2) Spectroscopic redshift of the object.
  (3) Logarithm of mass of the SMBH in solar masses, using
  the results in \citet{2011ApJS..194...45S}, the FWHM of
  \ion{Mg}{II} BEL (model S10).
  (4) Time baseline of our velocity shift measurement.
  (5) Velocity shift measured with BEL specified in column (3).
  (6) LoS acceleration.\\
  Spectra and correlation functions of 
  objects in this table, which have prominent $|\Delta V|$,
  are shown as examples in Fig. \ref{fig:spec_example}; 
  they are: J014822.62+132142.6, J125238.51+115557.1, and
  J163709.31+414030.8.
}
\label{tab:high-v-high-snr}
\end{deluxetable*}
% J212017.00+004841.7 is interesting--its Mg II window shows
% shifts on both emission and absorption features. This one
% is commented out in the table.

\begin{figure}
  \centering
  \includegraphics[width=3.1in, keepaspectratio]
  {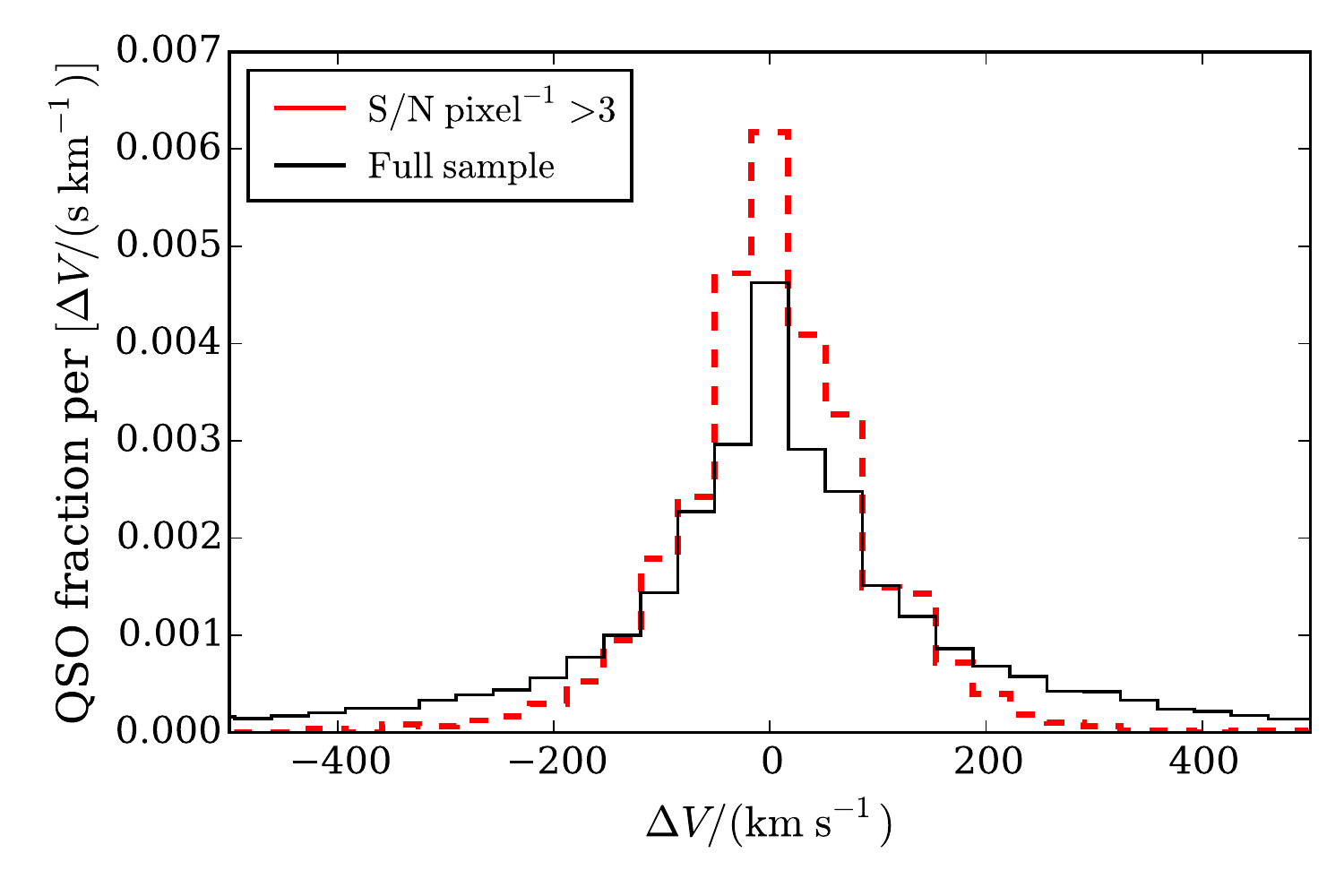}
  \caption{Distribution function of $\Delta V$, measured by
    cross-correlating the \ion{Mg}{II} BEL of the two epochs
    of observation, the latter coming from the SDSS
    DR12. The thin black histogram represents the
    distribution function of the whole sample, while the
    thick red dashed histogram is for the $\snr > 3$
    subsample. }
  \label{fig:dv_hist_snr3}
\end{figure}

In the SDSS DR12 sample of quasars with multiple
observations, we focus on the 1438 objects with $\snr > 3$
(section \ref{sec:sample-properties}) to investigate
$\Delta V$.  The distribution function of $\Delta V$, for
both the whole sample and the $\snr > 3$ subsample, is
presented in Figure \ref{fig:dv_hist_snr3}. The distribution
function of the entire sample is centered at
$-2.6~\km~\s^{-1}$ with $\sigma\simeq 220~\km~\s^{-1}$. For
the $\snr > 3$ subsample, the distribution function is
centered at $9.8~\km~\s^{-1}$, with
$\sigma \simeq 90~\km~\s^{-1}$.  For completeness, we
manually inspect those objects with intermediate $\snr > 1$
and large $|\Delta V| > 1000~\kms$. None of those
``high-$\Delta V$ outliers'' are trustworthy: the high
values result from highly discrepant pixels, or variable
absorption features. Therefore, in what follows, we focus on
the $\snr > 3$ sample.

We follow \citet{2013ApJ...777...44J} and tabulate all
targets with $|\Delta V| > 3\sigma \simeq 270~\km~\s^{-1}$
(that is, more than $3 \sigma$ outliers from the full
distribution) in Table \ref{tab:high-v-high-snr}. We have
removed 16 out of 33 objects where the apparent large
velocity was caused by highly discrepant pixels or broad and
variable absorption features, leaving 14 in the table. There
are very few objects that fall above our nominal
$|\Delta V|$ limit. Furthermore, we have already shown in
\S\ref{sec:follow-up} above that most of the velocity shifts
of order $\sim 300~\kms$ are not due to orbital
motion. However, we tabulate these quasars for completeness
and to demonstrate the small total number of them. In
addition, one (and the only) object in Table
\ref{tab:high-v-high-snr} with
$|\Delta V| > 5\sigma = 450~\km~\s^{-1}$ is presented in the
bottom row of Figure \ref{fig:spec_example}.

The percentage of objects showing outlying values of
$|\Delta V|$ is quite small. However, given that we are
probing typical timescales of a decade, we would expect to
encounter some significant velocity shifts from a
population of sub-pc binary BHs among typical QSOs.  In the
next section, we use gas-assisted inspiral models from
\citet{2013ApJ...774..144R, 2016ApJ...827..111R} to quantify
this expectation.

\section{Modeling and interpretation}
\label{sec:modeling}

To interpret our findings in terms of the SMBH binary
occurrence rates, we need to know the fraction of time that
a binary with a given set of characteristics (mass, mass
ratio, etc.) spends at a particular orbital separation
through its lifetime. That requires knowledge of the
residence time $\left|\d\ln a/\d t\right|^{-1}$, which can be
computed once the processes driving binary inspiral are
specified.

Following \citet{2013ApJ...777...44J}, we will
focus on the sub-pc orbital evolution of the binary driven
by its gravitational coupling with the surrounding
circumbinary disk, which must exist around the binary to
fuel its quasar activity. Tidal interaction of the binary
with the disk \citep{1980ApJ...241..425G,
  1994ApJ...421..651A} results in angular momentum exchange
between the two, leading to binary inspiral. The angular
momentum of the sub-parsec binary is absorbed by the gas in
the inner regions of these accretion disks, and is then
transported outwards by viscous stresses
\citep{2013ApJ...774..144R, 2016ApJ...827..111R}.

In this work we follow the description of the disk-binary
coupling in \citet{2013ApJ...774..144R}, which should be
consulted for details. In these models one self-consistently
follows the coupled viscous evolution of the disk and the
orbital inspiral of the binary. In particular, the non-local
nature of the viscous coupling (i.e. the time evolution of
the ``radius of influence'' out to which internal stresses
propagate binary perturbation at a given moment of time) is
fully accounted for, compared to other (quasi-) steady-state
models \citep[e.g.][]{2009ApJ...700.1952H} or purely
self-similar solutions
\citep[e.g.][]{1999MNRAS.307...79I}. The behavior of the
internal stress in our circumbinary disk models is
characterized by the dimensionless viscosity parameter
$\alpha = 0.1$.

One of the key characteristics that determines both the SMBH
inspiral and the disk evolution in the models of
\citet{2013ApJ...774..144R} is the gas accretion rate
through the disk far outside the binary orbit
$\dot{M}_\infty$. We often quantify this rate via the
Eddington ratio
$\dot{m}_\edd \equiv \dot{M}_\infty / \dot{M}_\edd$, which
is a free parameter in our calculations.  The value of
$\dot{M}_\edd$ is determined for the mass of the
secondary\footnote{We adopt radiative efficiency
  $\epsilon = 0.1$ as is commonly assumed for the AGN
  accretion disks.}, which is assumed to be the main
accretor (see below).  The disk-binary evolution models of
\citet{2013ApJ...774..144R} assume that a binary accretes
only a small fraction (e.g. below several tens of per cent)
of the incoming gas, in agreement with
\citet{2008ApJ...672...83M}.  Most of the mass accumulates
in the inner disk, increasing the rate of the angular
momentum exchange between the disk and the binary. As
demonstrated by \citet{2016ApJ...827..111R}, this assumption
results in the highest possible torque on the binary for a
given $\dot{M}_\infty$, meaning the {\it fastest} orbital
inspiral. We will comment on the implications of this
assumption in \S\ref{sec:discussion}.

Another important parameter of the model is the binary mass
ratio $q \equiv M_\s / M_\p$ --- the ratio of the secondary
to primary BH mass. It is generally thought that the
secondary BH should intercept most of the accretion flow
from the circumbinary disk
\citep[e.g.][]{2009MNRAS.393.1423C, 2011MNRAS.415.3033R,
  2012A&A...545A.127R, 2014MNRAS.439.3476R,
  2014ApJ...783..134F}.  If the secondary BH is 
luminous enough to be seen as a quasar at $z \approx 1$
(i.e., be included in our sample) than the mass ratio must
be close to unity, since the observed luminosities already
require secondary BH masses $>10^8~\msun$. For that reason,
in this work we explore the values of $q$ ranging from
$10^{-2}$ to $1$.

The binary typically starts at large separation in the so
called ``disk dominated regime'', when the ``local disk
mass'' $M_\d\equiv \Sigma r^2$ ($\Sigma$ is the disk surface
density) exceeds the mass of the secondary $M_\s$. Then the
binary inspiral is believed to be governed by the viscous
timescale of the disk, making the residence timescale
largely independent of the secondary mass:
$|d\ln a/dt|^{-1}\sim t_\nu\equiv r^2/\nu$.  However, later
on, as the binary orbit shrinks, the local disk mass becomes
lower than $M_s$, according to our calculations that fully
account for the back-reaction of the binary torque on the
circumbinary disk structure. This effect ultimately makes
the residence time sensitive to the mass of the secondary.

Our calculations assume that binary orbits are
circular. There are simulations showing that high
eccentricity may arise through binary-star interactions
\citep[e.g.][]{2008ApJ...686..432S, 2010ApJ...719..851S,
  2011ApJ...732L..26P, 2012ApJ...749..147K,
  2015MNRAS.454L..66S, 2015ApJ...810...49V} and binary-disk
interactions \citep[e.g.][]{2005ApJ...634..921A,
  2009MNRAS.393.1423C, 2011MNRAS.415.3033R,
  2014MNRAS.439.3476R}. However, given the lack of
observational evidence regarding this issue, we opted for
the simplest possible assumption of circular orbits.

We start the binary at $0.1~\pc$, which is a typical
stalling radius for the inspiral driven by purely stellar
dynamical processes \citep{2002MNRAS.331..935Y}. We also
consider models with $0.3~\pc$ initial separation, which are
used for comparison. The binary BH orbit is evolved until
its semi-major axis becomes equal to the radius of BLR
\citep[e.g.][]{2010ApJ...725..249S},
\begin{equation}
  \label{eq:size-blr}
  R_\mathrm{BLR} = 2.2\times 10^{-2}~\pc\times
  \left( \dfrac{L_\bol}{1.26\times 10^{45}~\erg\
      \s^{-1}} \right)^{3/2}~.
\end{equation}
At this point, we assume that the BLR around the secondary
is destroyed and we may no longer expect to detect periodic
shifts of the broad emission lines due to the orbital motion
of the secondary. This radius represents a large uncertainty
in our method.

\begin{figure}
  \centering
  \includegraphics[width=3.1in, keepaspectratio]
  {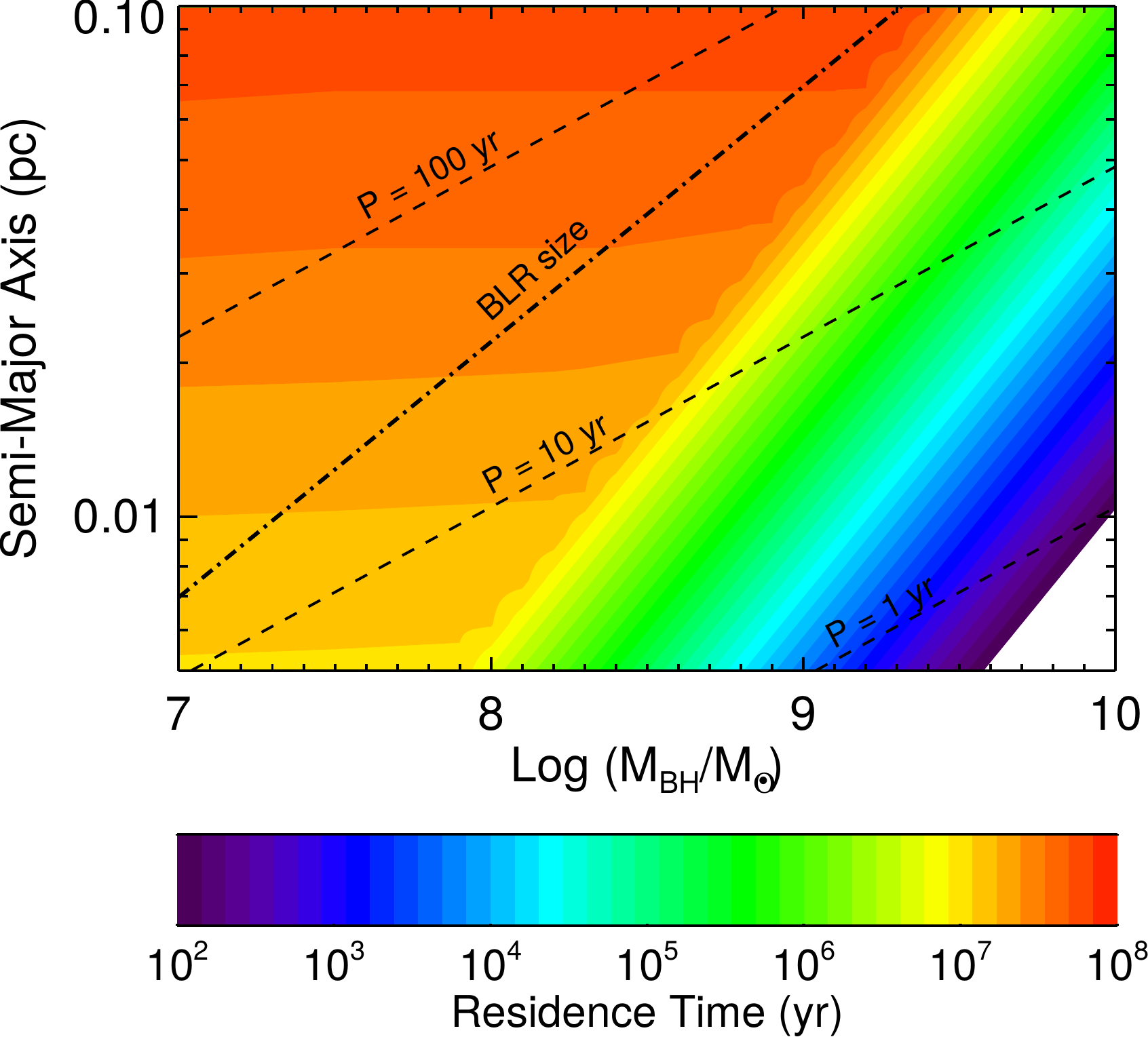}
  \caption{SMBH binary residence time $t_\mathrm{res}$, as a
    function of radius $r$ and the total mass of the binary
    system $M_\bh$. Here we only show the ``fiducial''
    example \citep[see others in][]{2013ApJ...777...44J},
    which has Eddington ratio $\dot{m}_\edd = 0.1$ and SMBH
    mass ratio $q = 1$. The dashed lines are the overlaid
    contours of constant orbital period. The dash-dotted
    line indicates the size of the BLR. 
    % {\bf I think we should go back to rainbow.}
    }
  \label{fig:tres_example}
\end{figure}

Using these assumptions, we calculate the residence time as
a function of radius as the binary evolves under the
influence of tidal coupling to the disk.  An example of the
calculated residence times is shown in Figure
\ref{fig:tres_example}, as adapted from
\cite{2013ApJ...777...44J}. The figure shows the fiducial
case, where the Eddington ratio $\dot{m}_\edd = 0.1$, and
mass ratio $q = 1$. As the accretion rate $\dot{m}_\edd$
increases, or the mass ratio $q$ decreases, disk torques
become more important to the motion of secondary, hence
shortening the residence time.  In practice, grid data of
separation $\{r_i\}$ and the corresponding evolution time
$\{t_i\}$ are generated, which will be utilized in the
following discussions.

We use the predicted residence times as a function of
orbital separation to calculate (a) what fraction of objects
would be observable with velocity shifts greater than
$450~\kms$, as a function of the total time baseline and (b)
the full expected distribution of velocity offsets. These
fractions assume that {\it all} the QSOs are binaries.  To
perform this calculation, we adopt a range of BH masses
based on luminosity (and assuming a range of accretion
rates, described below), and
$\Delta \tau \equiv \Delta t / (1 + z)$ (the time lapse in
the object rest frame) between the most widely separated
epochs. Once again, we always assume that only the secondary
BH carries a BLR.

With these assumptions, we obtain an interpolated grid of
data, giving the relation, on the $i$th grid point, between
the separation of the binary $r_i$ and the time $t_i$ that
the binary takes to evolve from $r=0.1~\pc$ to this
$r_i$. On each grid point, we calculate the expected value
of $|\Delta V|$, after marginalizing over orbital phase
($0<\phi<2\pi$) and the cosine of inclination angle. To be
most general, we consider that we could preferentially miss
nearly edge-on systems due to obscuration
\citep{1985ApJ...297..621A}. 
{\bf We parametrize the opening
angle of the obscuration with a critical angle $i_c$, such
that objects with $i_c<i<(\pi-i_c)$ are likely to be
obscured. We consider two cases, one with no obscuration,
$i_c=\pi/2$, and the other with $i_c=\pi/3$ \citep[e.g.][]
{2008A&A...490..905H}
\begin{equation}
  \label{eq:dv-single-r}
  \begin{split} \langle | \Delta V | \rangle_i & =
V_\mathrm{co}(r_i) \left\langle | \cos (\phi + \Omega \Delta
\tau ) - \cos \phi | \right\rangle_\phi \langle |\sin i|
\rangle_{\cos i} \\
% & = \dfrac{\pi}{4} V_\mathrm{co}(r_i)
% \left\langle 2 \left| \sin \left(\phi + \dfrac{\Omega\Delta
% \tau}{2} \right) \sin \left(\dfrac{\Omega\Delta
% \tau}{2}\right) \right| \right\rangle_\phi \\
& = V_\mathrm{co}(r_i) \sin \left(\dfrac{\Omega\Delta
    \tau}{2}\right)\times \left[ \dfrac{2 i_c - \sin (2
    i_c)}{\pi(1-\cos i_c)} \right]\ ,
  \end{split}
\end{equation} 
where $\Omega$ is the angular velocity of
orbital motion specified by equation \eqref{eq:omega_orbit}
at radius $r_i$, $V_\mathrm{co}(r_i)$ is the magnitude of
the orbital linear velocity at $r_i$ relative to the center
of mass, and $M_\bh$ is again the mass of the secondary
instead of the total,
\begin{equation}
  \begin{split} V_\mathrm{co}(r_i) & \equiv \left[ \dfrac{G
M_\bh} {r q ( 1 + q )} \right]^{1/2}\ .
   \end{split}
\end{equation} It is straightforward to prove that
eq. \eqref{eq:dv-single-r} reduces to $\langle | \Delta V |
\rangle_i = V_\mathrm{co}(r_i) \sin (\Omega\Delta \tau/2)$
at $i_c=\pi/2$, our fiducial case.}

For every $r_i$, a value of $\langle |\Delta V| \rangle$ is
evaluated, which is weighted with $(\delta \tau)_i \equiv
(\tau_{i+1} - \tau_{i-1})$ to generate a histogram. This
histogram is, by definition, the probability distribution
function of $|\Delta V|$ of this quasar. Stacking the
distribution functions of every quasar in our sub-sample, we
obtain the expected distribution function of $|\Delta V|$ of
the whole ensemble. The mass ratio $q$, accretion rate
$\dot{m}_\edd$ relative to the Eddington limit, and BLR size
$R_\mathrm{BLR}$ of these quasars are quite uncertain. To
bracket the resulting uncertainties, we assume different $q$
($1$ and $0.1$), $\dot{m}_\edd$ ($0.1$ and $1$), and
$R_\mathrm{BLR}$ [$0.5-2$ times the size in equation
\eqref{eq:size-blr}], to explore a plausible region in
parameter space.

\begin{figure} \centering
  \includegraphics[width=3.1in, keepaspectratio]
{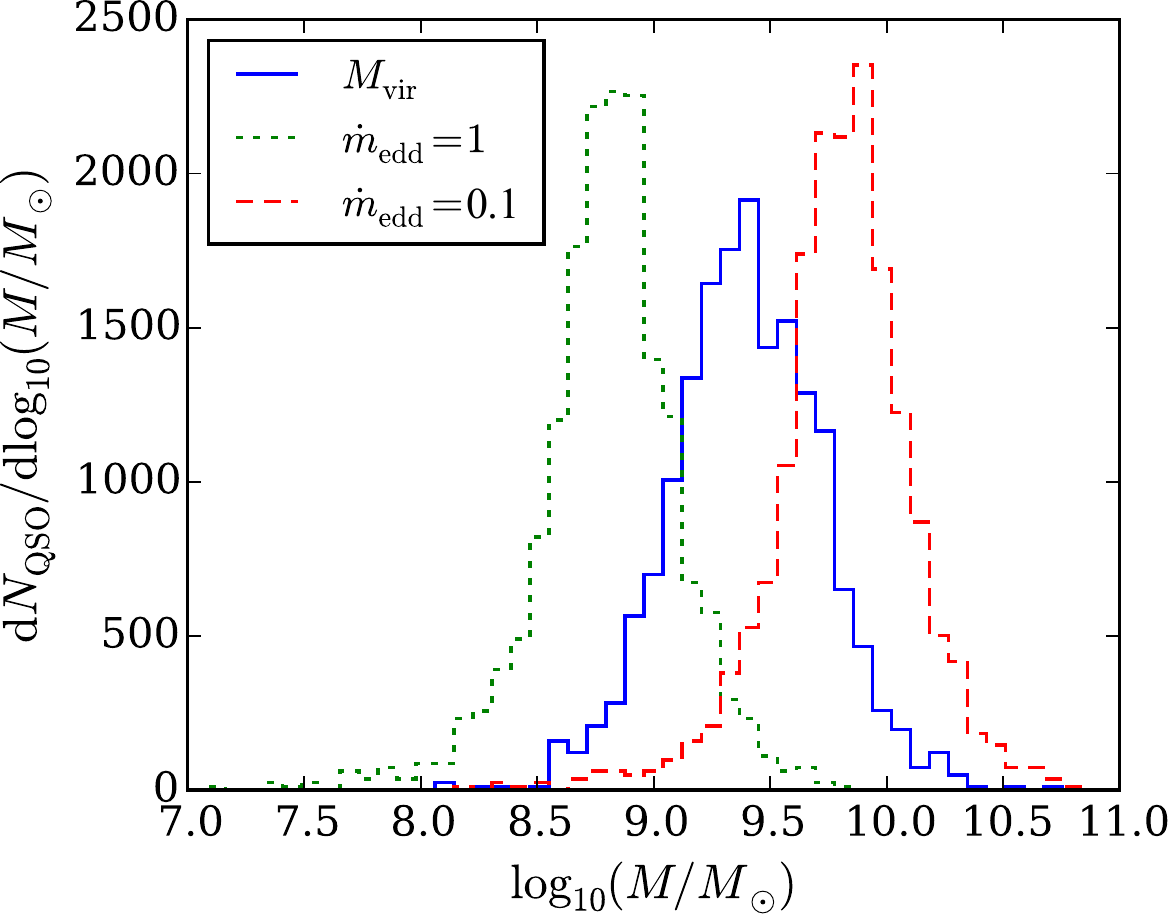}
  \caption{Secondary SMBH mass estimated by different
methods, using \ion{Mg}{II} line width \citep[blue
histogram][]{2011ApJS..194...45S}, and bolometric luminosity
assuming $\dot{m}_\edd = 0.1$ (red dashed histogram) or
$\dot{m}_\edd = 1$ (green dotted histogram). }
  \label{fig:mass_bracket}
\end{figure}

To make our calculations self-consistent, we adopt a single
luminosity for the BH as measured, and then assume that the
BH is radiating at either 10 or 100 per cent of the
Eddington luminosity to derive the BH mass.  These two
assumptions bracket the virial estimate based on
\ion{Mg}{II} from \citet{2011ApJS..194...45S}, as shown in
Figure \ref{fig:mass_bracket}.

\begin{figure*} \centering
  \includegraphics[width=6.3in, keepaspectratio]
{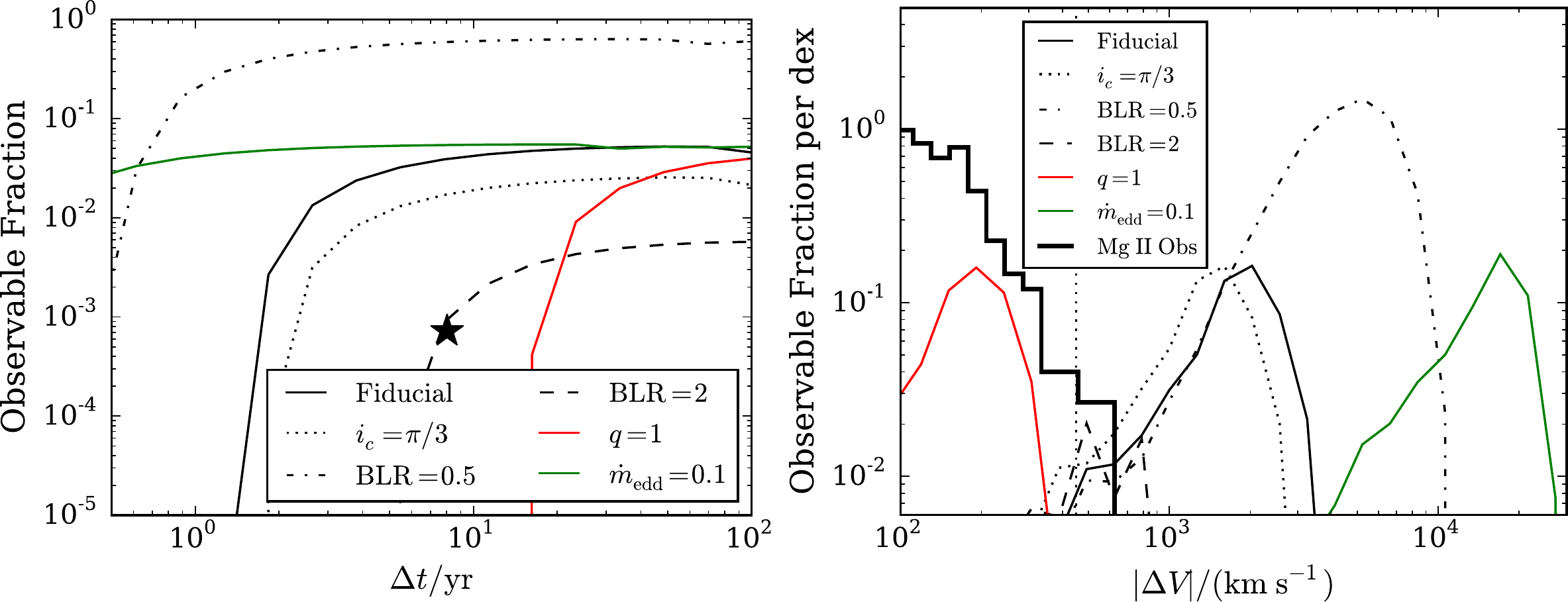}
  \caption{Observability of binary SMBHs, comparing
theoretical predictions and observation data.  Our
observation results are from the $\snr > 3$ sub-sample using
the SDSS DR12 observation data, measuring the \ion{Mg}{II}
BEL cross-correlation velocity shift.  {\bf Left panel}:
fraction of SMBHs that have greater velocity shift than the
$5\sigma$ ($450~\km~\s^{-1}$) threshold. Observation result
is indicated by the star symbol. The ``fiducial'' light
black solid curve has $q=0.1$, $\dot{m}_\edd=1$,
``standard'' BLR size (equation \ref{eq:size-blr}), and
$i_c=\pi/2$. Other curves have one parameter different from
the fiducial one, which is indicated in the legend.  {\bf
Right panel}: the distribution functions of
$\log_{10}|\Delta V|$. The thick black histogram on the left
shows observation results. Theoretical curves have the same
denotations with color and line style as in the central
panel.  A vertical dashed line shows the $|\Delta V| =
5\sigma \simeq 450~\km~\s^{-1}$ threshold.  }
  \label{fig:dv_observability}
\end{figure*}

Figure \ref{fig:dv_observability} presents the two
statistics (observable fraction and $|\Delta V|$).  In the
left panel, the upper limit on our observational fraction is
represented by the black star. The right panel shows the
expected distributions in velocity, again assuming that the
target binary starts evolving at $r=0.1~\pc$. The expected
velocity distributions range from $500$ to $10^4$
km~s$^{-1}$. As we discuss in \S \ref{sec:discussion}, given
the typical lifetimes of quasars and our estimated residence
times, we would naively expect to detect a few large
velocity shifts.  That we find no velocity shifts of this
magnitude places a real constraint on the length of time
that binary BHs spend at sub-pc separations in an active
phase, even given substantial uncertainties on the BLR size.

\section{Discussion and Summary}
\label{sec:discussion}

This paper presents two exercises aimed at constraining the
statistics of binary SMBHs with sub-pc separation. First, we
present follow up of the binary SMBH candidates first
presented in \citet{2013ApJ...777...44J}, using new
observations of 21 of the quasars that were observed to
display radial velocity shifts $>270~\km~\s^{-1}$ in that
paper.  Here we examine whether the observed radial velocity
shifts grow larger with longer time baselines, as would typically be
expected if they were due to orbital
motion. Cross-correlation of the \ion{Mg}{II} broad emission
line region over roughly a decade shows considerably lower
empirical line-of-sight acceleration ($a_\los$) in each
object as compared to the previous detections, and is
inconsistent with orbital motion.

Second, we extend the methodology of
\citet{2013ApJ...777...44J} and examine repeat observations
of these quasars in the SDSS/BOSS data. Among them, 1438
have a \ion{Mg}{II} BEL detected with $\snr > 3$. The
velocity shift distribution function, measured through
cross-correlation, peaks at $\sim 9~\km~\s^{-1}$ and has
$\sigma\sim 90~\km~\s^{-1}$ as its dispersion. The high S/N
subsample yields 15 objects with $|\Delta V| > 3\sigma
\simeq 270~\km~\s^{-1}$, while in the $|\Delta V| > 5\sigma
\simeq 450~\km~\s^{-1}$ regime there is only one
object. Since we have demonstrated that velocity shifts of
$\sim 300~\kms$ can occur in quasars without binaries, we
prefer to adopt the more conservative value of $450~\kms$ as
a possible indicator of orbital motion in a
binary. Comparing this single candidate with theoretical
expectations, we conservatively find that $\lesssim 1.6$ per cent of
SMBH reside in sub-pc binaries\footnotetext{We also tested
increasing the sample by adopting a lower SNR limit ($\snr >
1$). We derive a similar limit on the sub-pc fraction with
that alternate sample.}.
  
Our simulations suggest that quasars spend at least a Myr to
reach radii where gravitational radiation can bring
coalescence.  Given fiducial quasar lifetime estimates of
$10^7-10^8$ yrs \citep{2001ApJ...547...12M}, we should have
detected a few binaries in our sample.  This inferred ratio
is quite sensitive to the uncertainties in the BH mass
measurements and BLR sizes.  Nevertheless, given our
relatively long time baselines and large numbers of objects,
we should be detecting large velocity shifts if merging BH
binaries spend the amount of time expected from simulations
in a sub-pc quasar phase.

Assuming fiducial values for the BH binary of $\dot{m}_\edd
= 0.1$ and $r_0=0.1~\pc$, and a BLR size in accordance with
\citealt{2010ApJ...725..249S}, we find that $\lesssim 1$ per
cent of the quasar population is in a sub-pc phase. These
are the only observational limits on the residence times of
BH binaries with $\sim 0.1$ pc separations at cosmological
distances, and (if taken at face value) they suggest that BH
binaries spend a surprisingly short period of time in this
phase compared to our models.

There are some major caveats. As is shown in Figure
\ref{fig:dv_observability}, the upper limit on the sub-pc
binary fraction is extremely sensitive to BLR size, from 0.1
per cent with a BLR size one half as large, to $\sim 100$
per cent with a doubled BLR size. If the starting point is
set to be as large as $0.3~\pc$, our limit on the observable
fraction drops to $\lesssim 0.2$ per cent.

On the other hand, in our gas-assisted binary inspiral
scenario, the binary occurrence rate should be independent
of the accretion history of the binary. Indeed, we observe
the binary as an AGN only while it is actively accreting,
and this is also the period when its orbit is evolving due
to disk torques. As long as the mass supply switches off and
the gas-driven binary inspiral stalls (assuming that the
gravitational wave emission is not yet capable of shrinking
the orbit efficiently) we simply stop observing such
systems, which thus naturally removes them from our sample
of observed objects and does not affect the statistics of
luminous AGNs.

It should also be kept in mind that our model assumptions
tend to {\it minimize} the estimated BH binary residence
time for a given separation. As mentioned in
\S\ref{sec:modeling} our assumption of significantly
suppressed accretion onto the binary (compared to $\dot
M_\infty$) results in the highest value of the angular
momentum loss experienced by the binary (for a given $\dot
M_\infty$), meaning that it spends the shortest time per
interval of the semi-major axis. Allowing the binary to
accrete from the disk more readily, in agreement with some
recent simulations \citep{2014ApJ...783..134F,
2015ApJ...807..131S} would reduce mass accumulation in the
inner disk and {\it lower} the torque on the binary, as
demonstrated in
\citet{2013ApJ...774..144R,2016ApJ...827..111R}.  As a
result, the binary would spend {\it more time} in the
separation range probed by the cadence of our sample,
exacerbating the conflict with our non-detection of
significant line shifts. Thus, our model assumptions
effectively imply that the inferred constraint on the SMBH
binary occurrence rate is in fact an {\it upper limit}.  In
that sense, our conclusion that this phase of binary
evolution should be shorter than expected is robust.

First, we address the origin of the $\sim 300~\kms$ %
velocity shifts that we observe in a small sub-sample of %
objects.  The \citet{2013ApJ...777...44J} sample display %
large velocity shifts that cannot be ascribed to orbital %
motion.  We have asserted that the $\sim 300~\kms$ %
velocity shifts are not due to noise. Here, we address the %
origin.  Other possible sources must be considered to %
explain the velocity shift. One possibility is recoiling %
SMBHs.  These might well present similar effects in terms %
of velocity offset, but with temporal changes that are not %
sustained in a way similar to binary systems %
\citep[e.g.][] {1973ApJ...183..657B, 1984MNRAS.211..933F, %
2005ApJ...635..508B, 2010ApJ...717..209C, %
2012ApJ...752...49C, 2013MNRAS.428.1341B}.

We can also compare our conclusions with results from other
techniques to find the electromagnetic signatures of tight
binaries.  These techniques often probe shorter orbital
periods, but also point towards short residence times. In
particular, using quasi-periodic photometric variability,
\citet{2015MNRAS.453.1562G} present 111 candidate binaries
out of a sample of 243,500 quasars mostly with $z\sim 1-2$
but extending to $z=4$.  Similarly,
\citet{2016arXiv160401020C} find 33 candidates out of 35,383
quasars with a median redshift $z\sim 2$, and
\citet{2016arXiv160909503L} identify $<1$ candidates out of
670 quasars.
%{\bf add http://adsabs.harvard.edu/abs/2016arXiv160909503L}
Given their orbital periods of 0.3-6 yrs, 
these candidates have orbital
separations a factor of 7-40 smaller than ours, orbital
separations that we are not sensitive to with our
method. Also, at these radii the inspiral may be dominated
by gravitational radiation.  According to
\citet{2015MNRAS.453.1562G}, their detected fraction of
$10^{-4}$ is a factor of $\sim 5$ lower than expected based
on simulations \citep*{2009ApJ...703L..86V}.

% Of course, these binary candidates would be at yet smaller
% separations where the residence times are shorter.

Our finding (assuming that $\dot{m}_\edd = 0.1$ and %
$r_0=0.1~\pc$, and a BLR size in accordance with %
\citealt{2010ApJ...725..249S}) that $\lesssim 1$ per cent %
of the quasar population is in a sub-pc phase is %
consistent with other existing observational limits. When %
we adopt an initial radius of $r_0=0.3~\pc$, the %
population fraction drops to 0.3 per cent.
 
At lower redshifts, based on double-peaked broad emission
lines, \citet{2009Natur.458...53B} find that only
$10^{-4}$-$10^{-3}$ quasars are in a sub-pc phase (similar
to our raw limit of a few $\times 10^{-3}$).  Nominally,
these results are also consistent with the sub-pc binary
fractions found by \citet{2009ApJ...703L..86V}, at
comparable redshifts of $z \lesssim 0.7$. In the context of
their merger-driven quasar model, they report that the
observable fraction of binary BHs would increase by a factor
of $5-10$ if they examined $0.7 < z < 1$. Simulations tuned
to our redshift window are needed to determine the expected
sub-pc binary fraction, given quasars triggered by merging
and gas-assisted merging timescales \citep[e.g.][]
{2016MNRAS.tmp.1488K}.
% {\bf cite Kelley et al here}

The Pulsar Timing Arrays (PTAs) are also gaining the
required sensitivity to rule out regions of parameter space
for tight SMBH binaries \citep[e.g.,][]{2016ApJ...819L...6T,
2016arXiv160206301T}. Pulsar timing is sensitive to the
stochastic superposition of weak gravitational waves from
SMBH binaries at sub-parsec scales, at which SMBH binaries
are supposed to spend some of their life time.
\citet{2015Sci...349.1522S} were able to rule out the most
optimistic theoretically predicted numbers of sub-pc BH
binaries (based on contemporary star-assisted or
gas-assisted dissipation models,
e.g. \citealt{2003ApJ...590..691W},
\citealt{2015ApJ...799..178K}, and especially
\citealt{2015MNRAS.447.2772R}) using existing PTA
constraints on the stochastic gravitational wave
background. It is interesting that our constraints and those
from the PTAs point in a similar direction towards short
residence times at small separations.

The low frequency of the BH binary systems, and in
particular the lack of detection of high-velocity line
shifts, start to place an interesting limit on the binary
SMBH evolution picture. Although our current interpretation
of the observations uses a particular model of binary
inspiral, we still believe that it provides an interesting
constraint.

If all quasars are triggered by merging, and the AGN
lifetime is only $\sim 10^8~\yr$, then our results indicate
that quasars spend a much lower fraction of their lifetimes
in a sub-parsec phase than our relatively conservative
gas-assisted theories predict.  Given the limitations of our
method, we are left with a few possible interpretations of
our findings.  One is that the binary BH (or at least the
secondary) does not radiate efficiently in this phase.
Second, the BLR may be much larger than we believe,
encompassing the whole binary, such that the signs of the
orbital motion cannot be observed. Third, all BH binaries
may stall at radii larger than $\gtrsim 0.3~\pc$, outside of
our range of interest.  Fourth, the most interesting
possibility is that BH binaries merge much more rapidly than
our models predict.

Extensions of our work would involve detections of velocity
shifts using more broad emission lines. For example, the
\ion{C}{IV} $\lambda 1549~\ang$ BEL is often more luminous
than \ion{Mg}{II} adopted in this paper. Nevertheless, the
\ion{C}{IV} BEL is generally expected to emerge from the
inner region of the BLR. Its characteristics are more
vulnerable to complicated quasar physics, especially
outflows, which are not fully understood at this moment
\citep[e.g.][]{2000ApJ...543..686P, 2004ApJ...616..688P,
2005ApJ...630L...9P, 2007ApJ...661..693P,
2014ApJ...789...19H}, showing absorption and asymmetry that
may introduce an extra bias into velocity shift
detection. Going forward, we must pursue as many independent
electro-magnetic constraints on the sub-pc binary BH
population as possible, as we prepare for the the first PTA
detections.  Our observational results also strongly
motivate further theoretical work to better understand the
orbital evolution of the SMBH binaries.

\acknowledgements
% \section*{Acknowledgments}

Roman R. Rafikov is an IBM Einstein Fellow at the Institute
of Advanced Study; his financial support is provided by the
NSF via grants AST-1409524, AST-1515763, NASA via grants
14-ATP14-0059, 15-XRP15-2-0139, and The Ambrose Monell
Foundation. Jenny E. Greene acknowledges funding from 
the National Science Foundation under Grant 
No. AAG: \#1310405.
We are thankful to Dr. Adam Myers, Dr. Michael Eracleous,
and Dr. Zoltan Haiman for useful comments and suggestions.

\bibliographystyle{apj}
\bibliography{binary_bh}

\end{document}